\documentclass[printer]{aa}

\usepackage{amsmath}
\usepackage{amssymb}
\usepackage{empheq}
\usepackage{mathrsfs}
				
\usepackage{amssymb}

\begin{document}

\title{A Self-Consistent Model for Dust-Gas Coupling in Protoplanetary Disks}

\author{Konstantin Batygin\inst{\ref{inst1}} \and Alessandro Morbidelli\inst{\ref{inst2}}}

\institute{Division of Geological and Planetary Sciences, California Institute of Technology, Pasadena, CA 91125, USA \email{kbatygin@gps.caltech.edu}\label{inst1} \and Laboratoire Lagrange, Universit\'e C\^ote d'Azur, Observatoire de la C\^ote d'Azur, CNRS, CS 34229, F-06304 Nice, France \email{alessandro.morbidelli@oca.eu}\label{inst2}}

                              
\abstract{Various physical processes that ensue within protoplanetary disks -- including vertical settling of icy/rocky grains, radial drift of solids, planetesimal formation, as well as planetary accretion itself -- are facilitated by hydrodynamic interactions between H/He gas and high-$Z$ dust. The Stokes number, which quantifies the strength of dust-gas coupling, thus plays a central role in protoplanetary disk evolution, and its poor determination constitutes an important source of uncertainty within the theory of planet formation. In this work, we present a simple model for dust-gas coupling, and demonstrate that for a specified combination of the nebular accretion rate, $\dot{M}$, and turbulence parameter, $\alpha$, the radial profile of the Stokes number can be calculated uniquely. Our model indicates that the Stokes number grows sub-linearly with orbital radius, but increases dramatically across the water-ice line. For fiducial protoplanetary disk parameters of $\dot{M}=10^{-8}\,M_{\odot}/$year and $\alpha=10^{-3}$, our theory yields characteristic values of the Stokes number on the order of $\mathrm{St}\sim10^{-4}$ (corresponding to $\sim$mm-sized silicate dust) in the inner nebula and $\mathrm{St}\sim10^{-1}$ (corresponding to $\sim$few-cm-sized icy grains), in the outer regions of the disk. Accordingly, solids are expected to settle into a thin sub-disk at large stellocentric distances, while remaining vertically well-mixed inside the ice line.}

\titlerunning{Feelin' Stoked}
\authorrunning{Batygin \& Morbidelli}

\maketitle

\section{Introduction}

All planets originate as dust, embedded within protoplanetary disks. As such, the physical properties and radial distribution of solid material within circumstellar nebulae play a key role in determining the efficiency of planetesimal formation \citep{YoudinGoodman2005,JY2007}, the principal mode of planetary accretion \citep{OK2010, LJ12}, as well as the large-scale architecture of the emergent planetary system \citep{Kokubo2012,Bitsch2019}. In other words, the overall process of planetary conglomeration is inherently controlled by the interactions between dust particles and the Hydrogen-Helium fluid that comprises the dominant component of the disk.

\begin{figure*}[h!]
\centering
\includegraphics[width=0.85\textwidth]{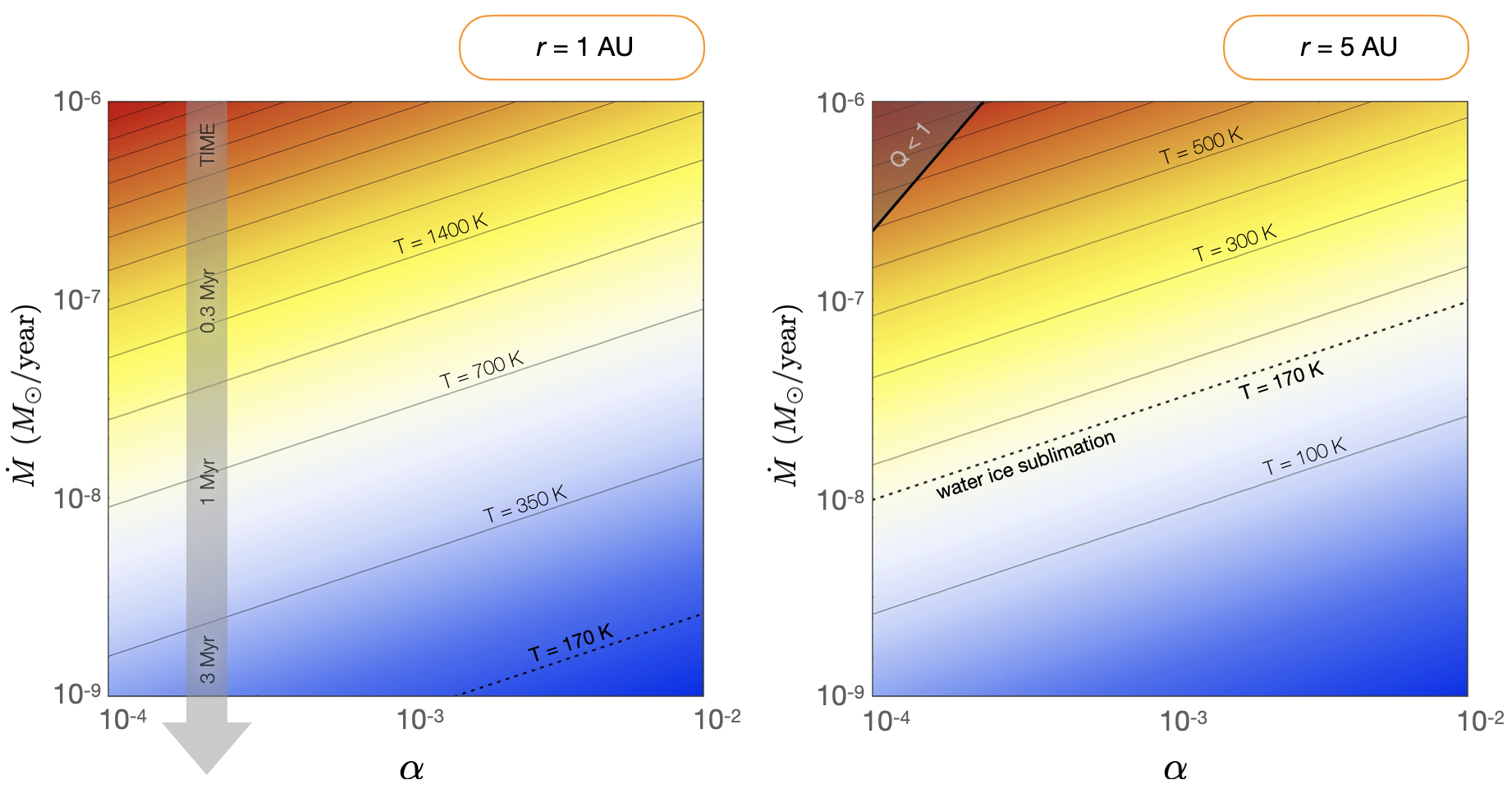}
\caption{Mid-plane temperature within the model (active) accretion disk at orbital radii of $r=1\,$AU (left panel) and $r=5\,$AU (right panel). Contours are shown on the $\dot{M}-\alpha$ plane, spanning the range of parameters relevant to young protoplanetary nebulae. Level curves corresponding to the assumed water ice sublimation temperature of $T_{\rm{ice}}=170\,$K are shown with dashed lines. As disks evolve, their accretion rates diminish, meaning that $\dot{M}$ constitutes a proxy for time, as shown on the left panel. A small region of gravitationally unstable parameter space where Toomre $Q<1$ is labeled on the top-left corner of the right panel.}
\label{fig:fig1}
\end{figure*}

The degree of coupling between dust and gas is quantified by a dimensionless quantity -- the \textit{Stokes number}\footnote{Defined as the product of the frictional timescale and the orbital frequency, the Stokes number is also referred to in the literature as the \textit{dimensionless stopping time}.}. In the context of astrophysical disks, the Stokes number specifies the fraction of an orbit that is required for a grain to re-equilibrate with the (sub-Keplerian) flow of the gas (e.g., \citealt{Adachi1976, Weiden1977}). Because the Stokes number depends both upon local disk properties and the particle size, disparate approaches to computing its value lead to considerable (and model-specific) degeneracies (\citealt{ChiangYoudin2010, 2019A&A...627A..83L,2020MNRAS.494.5134L} and the references therein). Therefore, establishing a connection between the Stokes number and local properties of the nebula is key to alleviating an important source of uncertainty in the theory of planet formation.

In the following text, we demonstrate that the Stokes number attainable by solid particles cannot be divorced from the disk model. We outline a simple theoretical model, and self-consistently compute the radial profile of the Stokes number within young protoplanetary nebulae. The compositional dependence of dust fragmentation threshold -- and by extension, the water-ice sublimation line -- plays a key role in our analysis. Informed by laboratory experiments, we assume that icy grains are more sticky than refractory particles. This distinction translates into corresponding fragmentation velocities that differ by an order of magnitude, and facilitates the main difference in the strength of dust-gas coupling between the silicate-rich inner disk and the ice-dominated outer disk.

We begin our calculation by recalling the basic properties of accretion disks in section \ref{sec:diskmodel}. We compute the Stokes number itself in section \ref{sec:Stnum}. Direct implications of our model are presented in section \ref{sec:res}. We summarize and discuss our results in section \ref{sec:conc}.

\section{Disk Model} \label{sec:diskmodel}

An essential property of protoplanetary disks is that they accrete onto their host stars. At any orbital radius, $r$, the relationship between the mass-accretion rate, $\dot{M}$ and the disk surface density, $\Sigma$, is given by (e.g., \citealt{Armitagebook}):
\begin{align}
\dot{M}=|2\,\pi\,r\,v_{r}\,\Sigma|.
\end{align}
A conventional assumption of a steady-state disk is obtained by supposing that $\dot{M}$ is independent of $r$. Within the framework of the viscous accretion paradigm \citep{LBP1974}, the radial velocity of the gas, $v_{r}$, arises from subsonic turbulence within the system. Adopting the conventional \citep{1973A&A....24..337S} prescription for turbulent viscosity, $\nu = \alpha\,c_{\rm{s}}\,h$, we have:
\begin{align}
v_{r}=-\frac{3}{2}\frac{\nu}{r}=-\frac{3\,\alpha\,k_{\rm{b}}\,T}{2\,\mu\,r\,\Omega},
\end{align}
where $\mu$ is the mean molecular weight, $k_{\rm{b}}$ is the Boltzmann constant, and $\Omega$ is the orbital frequency. Notice that in the above expression we have used the standard relation between the speed of sound and temperature $c_{\rm{s}}=\sqrt{k_{\rm{b}}\,T/\mu}$, and have expressed the disk scale height as $h=c_{\rm{s}}/\Omega$.

\begin{figure*}[h!]
\centering
\includegraphics[width=0.85\textwidth]{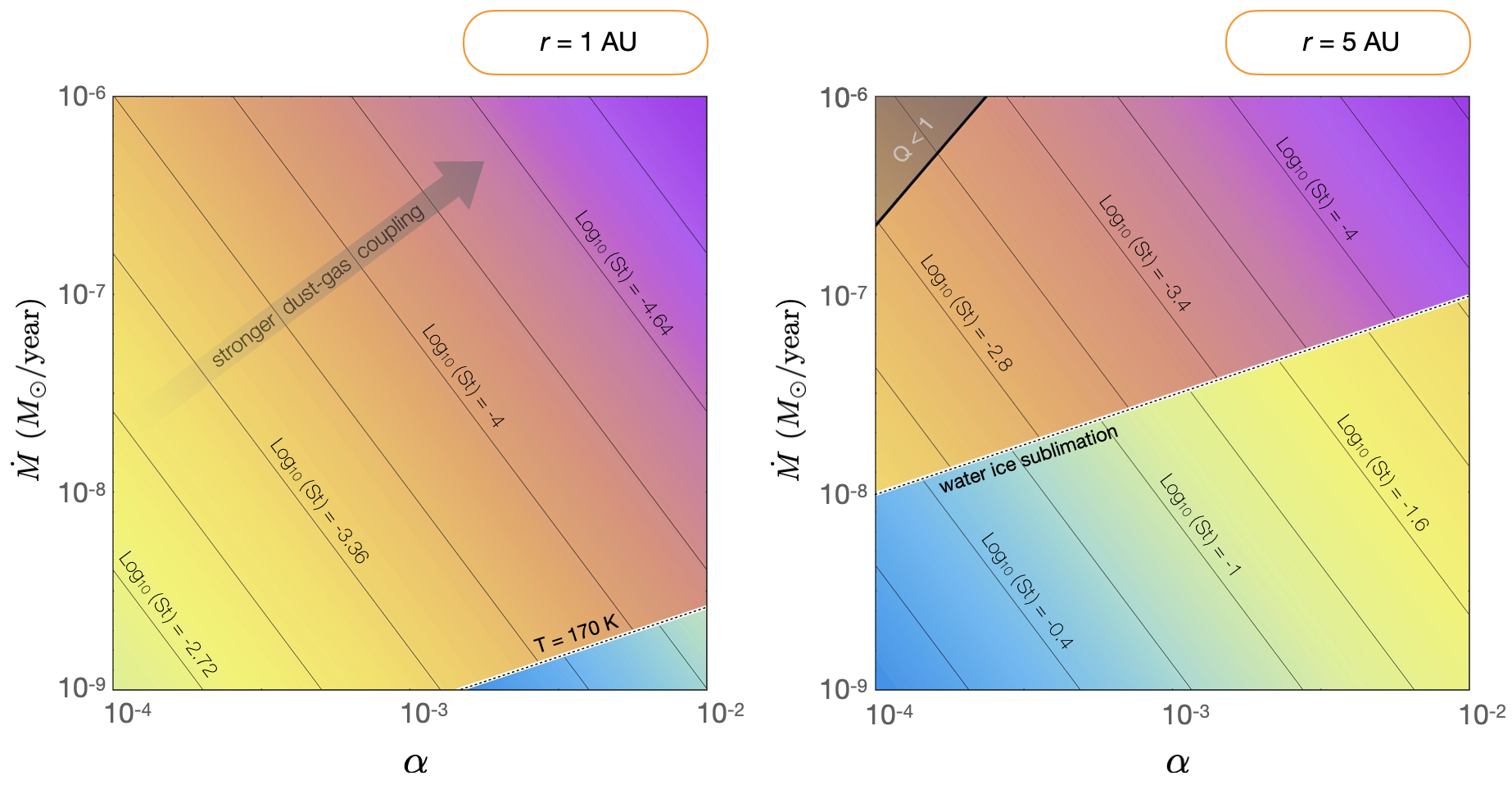}
\caption{Stokes number attainable by dust grains at $r=1\,$AU (left panel) and $r=5\,$AU (right panel). Owing to the fragmentation barrier, solid grains within the disk cannot grow to an arbitrarily large size, and disk turbulence acts as a limiting mechanism for particle growth. The inner regions of the disk are characterized by Stokes numbers on the order of $\rm{St}\sim10^{-4}$, indicating very tight gas-dust coupling. Conversely, icy grains within the outer disk can reach Stokes numbers of order $\rm{St}\sim10^{-1}$, implying diminished aerodynamic drag.}
\label{fig:fig2}
\end{figure*} 

The functional form of the mid-plane temperature of an actively heated optically thick disk stems from balance between viscous heat generation within the nebula and radiative losses (e.g., \citealt{Lee1991}):
\begin{align}
\sigma_{\rm{sb}}\,T^4=\frac{9\,\dot{M}\,\tau\,\Omega^2}{32\,\pi},
\end{align}
where $\sigma_{\rm{sb}}$ is the Stefan-Boltzmann constant, and $\tau$ is the optical depth. Assuming that dust acts as the dominant source of opacity, the optical depth can be expressed as \citep{Bitsch2014}:
\begin{align}
\tau=\frac{1}{2}\,f_{\rm{dust}}\,\Sigma\,\kappa_{\rm{dust}},
\end{align}
where $f_{\rm{dust}}\approx 0.01$ is the dust-to-gas ratio and $\kappa_{\rm{dust}}\approx30\,$m$^2$/kg (for simplicity, here we neglect the temperature-dependence; \citealt{BellLin1994}).

Combining these relations, we obtain an expression for the mid-plane temperature profile of the disk as a function of $\dot{M}$ and $\alpha$:
\begin{align}
T= \bigg( \frac{3\,f_{\rm{dust}}\,\kappa_{\rm{dust}}\,\mu\,\Omega^3\,\dot{M}^2}{64\,\pi^2\,k_{\rm{b}}\,\sigma_{\rm{sb}}\,\alpha} \bigg)^{1/5}.
\end{align}
As an illustrative example, the left and right panels of Figure (\ref{fig:fig1}) show level curves of the mid-plane temperature on the $\alpha-\dot{M}$ plane, at $r=1\,$AU and $5\,$AU, respectively. Additionally, on the left panel of the Figure, $\dot{M}$ is linked with disk age using the $\dot{M}-$age relationship presented by \citet{Hartmann1998}. Because the mid-plane temperature determines the speed of sound and the pressure scale-height of the disk, a given combination of $\alpha$ and $\dot{M}$ also uniquely specifies the profile of the disk viscosity.

\section{Stokes Number} \label{sec:Stnum}

The physical interpretation of the Shakura-Sunayev $\alpha$ is nothing more than the square of the turbulent Mach number (or equivalently, the ratio of stress to pressure). Correspondingly, the characteristic velocity scale at which turbulent eddies manifest within the nebula is $v_{\rm{turb}}\sim\sqrt{\alpha}\,c_{\rm{s}}$. Turbulent fluctuations within the gas give rise to a dispersion of velocities among the dust grains, such that particles that are coupled to the gas (i.e., those with Stokes number, $\rm{St}$, of unity or smaller) will experience collisions with velocities that are approximately given by \citep{Birnstiel2009}:
\begin{align}
\sigma_{\rm{dust}}\sim\sqrt{\rm{St}}\,v_{\rm{turb}}=\sqrt{\alpha\,\rm{St}}\,c_{\rm{s}}.
\label{sigmadust}
\end{align}
Note that herein, we have assumed that turbulent diffusivity of the gaseous and solid components of the disk are equivalent. In other words, our analysis is done for a turbulent Schmidt number of $\mathrm{Sc}=1$. Generalization of our results to any other value of the Schmidt number can be done by introducing a second turbulence parameter $\alpha'=\alpha/\mathrm{Sc}$ pertinent to the dust sub-disk.

A generic barrier to particle growth within the disk is presented by fragmentation, and experimental data indicate a fragmentation velocity of $v_{\rm{f}}^{\rm{rock}}\approx1\,$m/s and $v_{\rm{f}}^{\rm{ice}}\approx10\,$m/s and for silicate and water-ice particles, respectively\footnote{We note that recent literature points to an additional temperature-dependence of $v_{\rm{f}}$ for $T \ll T_{\rm{ice}}$ and for $T\sim T_{\rm{Si}}$ \citep{2019ApJ...873...58M, 2021A&A...652A.106P}, which we neglect for simplicity.} \citep{Blum1993,Stewart2009}. An equivalence between the fragmentation speed and the dust velocity dispersion thus yields a maximal Stokes number that grains can achieve within the nebula:
\begin{align}
\mathrm{St}_{\rm{max}}=\frac{1}{3}\frac{v_{\rm{f}}^2}{\alpha\,c_{\rm{s}}^2},
\end{align}
where the numerical factor of $1/3$ arises from a somewhat more detailed analysis \citep{OrmelCuzzi2007,Birnstiel2012}. Of course, in a realistic disk, dust particles follow a distribution of sizes, which translates to a range of relevant Stokes numbers. Nevertheless, the relatively rapid rate of growth of solid grains within the nebula\footnote{Particle growth timescale is well-approximated by $\tau_{\rm{grow}}\sim1/(\mathcal{Z}\,\Omega)$, where $\mathcal{Z}$ is the metallicity; \citep{2016A&A...594A.105D}.} is likely to ensure that the characteristic value of $\mathrm{St}$ is close to its maximal threshold. Consequently for the remainder of the paper, we assume that $\mathrm{St}\sim\mathrm{St}_{\rm{max}}.$

Substituting results from the previous section, we obtain the Stokes number as a function of $\alpha$ and $\dot{M}$:
\begin{align}
\mathrm{St}&=\frac{2\,v_{\rm{f}}^2}{3}\, \bigg(\frac{2\,\pi^2\,\sigma_{\rm{sb}}\,\mu^4}{3\,k_{\rm{b}}^4\,f_{\rm{dust}}\,\kappa_{\rm{dust}}\,\dot{M}^2\,\alpha^4\,\Omega^3} \bigg)^{1/5}.
\label{Stokesresult}
\end{align}
Drawing a parallel with Figure (\ref{fig:fig1}), in Figure (\ref{fig:fig2}) we project contours of the Stokes number onto the $\alpha-\dot{M}$ plane at $r=1\,$AU and $r=5\,$AU, assuming a H$_2$O condensation temperature of $T_{\rm{ice}}=170\,$K. Notice that for equivalent (and characteristic) values of viscosity and accretion rate (e.g., $\alpha\sim10^{-3}$, $\dot{M}\sim10^{-8}\,M_{\odot}/$year), at $r=1\,$AU and $r=5\,$AU we obtain $\rm{St}\sim2\times10^{-4}$ and $\rm{St}\sim10^{-1}$ respectively, implying very strong dust-gas coupling in the inner disk but only marginal coupling in the outer disk.

\begin{figure}[t!]
\centering
\includegraphics[width=\columnwidth]{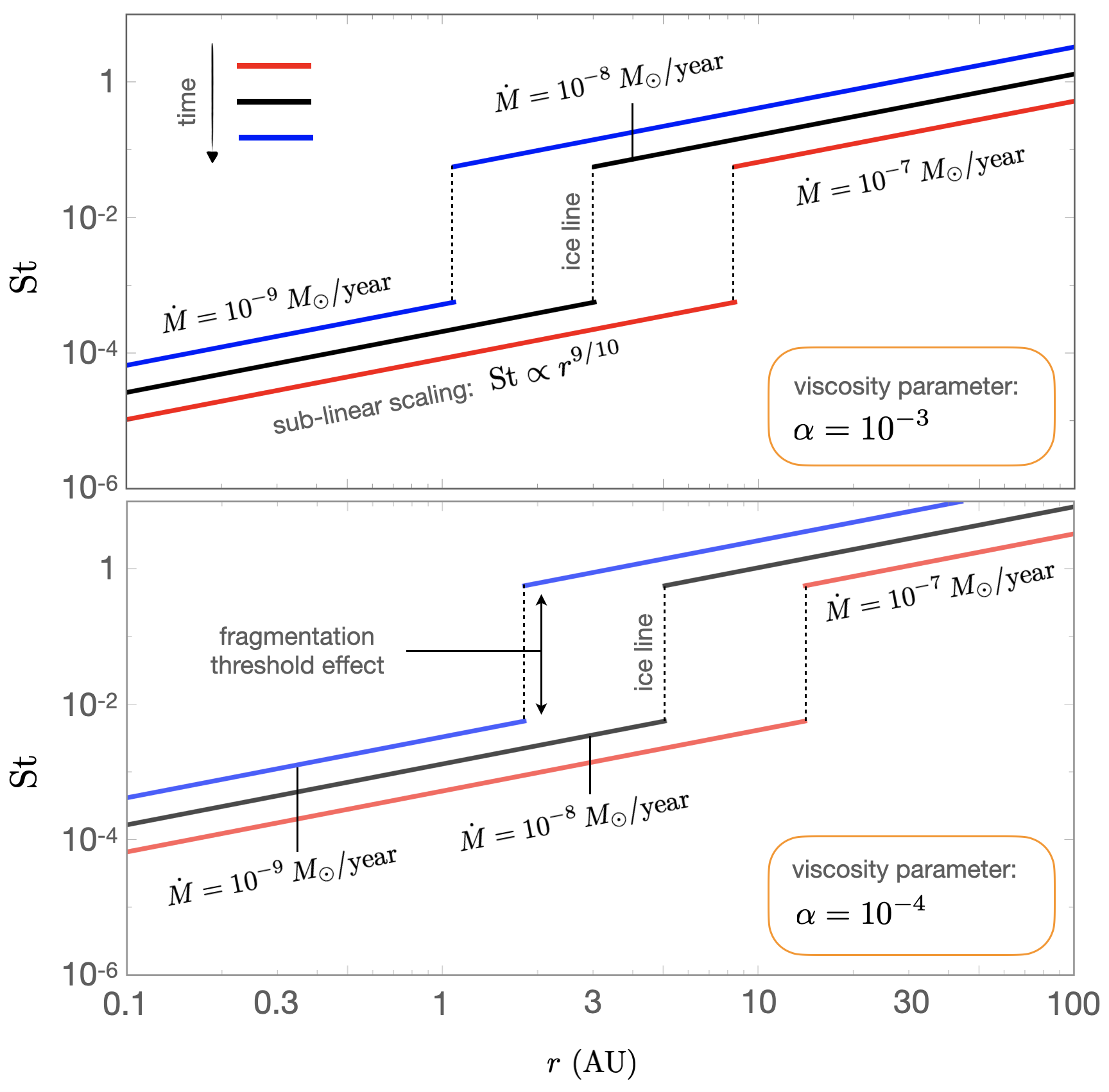}
\caption{Stokes number as a function of the orbital radius for Shakura-Sunayev $\alpha=10^{-3}$ (top panel) and $\alpha=10^{-4}$ (bottom panel). Both panels show curves corresponding to disk accretion rates of $\dot{M}=10^{-7}\,M_{\odot}/$year (red), $10^{-8}\,M_{\odot}/$year (black) and $10^{-9}\,M_{\odot}/$year (blue). Although quantitatively distinct, each profile is characterized by a $\mathrm{St}\propto r^{9/10}$ scaling, as well as a dramatic (factor of $\sim100$) increase in $\mathrm{St}$ across the ice sublimation front, owing to a change in the fragmentation threshold.} 
\label{fig:fig3}
\end{figure}

The overall functional form of equation (\ref{Stokesresult}) indicates that the Stokes number scales as the 9/10th power of the orbital radius and therefore increases in a marginally sub-linear manner with stellocentric distance. Moreover, the results depicted in Figure (\ref{fig:fig2}) highlight the importance of the water-ice line as a transition point within the disk, beyond which the value of the Stokes number increases dramatically due to the increase in $v_{\rm{f}}$. To illustrate the dependence of $\mathrm{St}$ on disk parameters, in Figure (\ref{fig:fig3}), we show $\rm{St}$ as a function of $r$ for various accretion rates and disk viscosity parameters of $\alpha=10^{-3}$ (top panel) and $\alpha=10^{-4}$ (bottom panel).

\section{Results} \label{sec:res}

A number of immediate implications follow from equation (\ref{Stokesresult}), and in this section we extend our model to compute the efficiency of vertical settling as well as the rate of radial drift of the dust. Additionally, we convert our estimates of $\rm{St}$ into a distribution of particle sizes within a protoplanetary disk and use our result to delineate the characteristic regimes of pebble accretion as a function of orbital radius as well as planet-to-star mass ratio. For the following calculations, we adopt a nominal combination of turbulence and accretion-rate parameters of $\alpha=10^{-3}$ and $\dot{M}\sim10^{-8}\,M_{\odot}/$year, which yield a model disk that is somewhat less massive than (but nevertheless comparable to) the \citet{Hayashi1981} minimum mass solar nebula.

\begin{figure}[t!]
\centering
\includegraphics[width=\columnwidth]{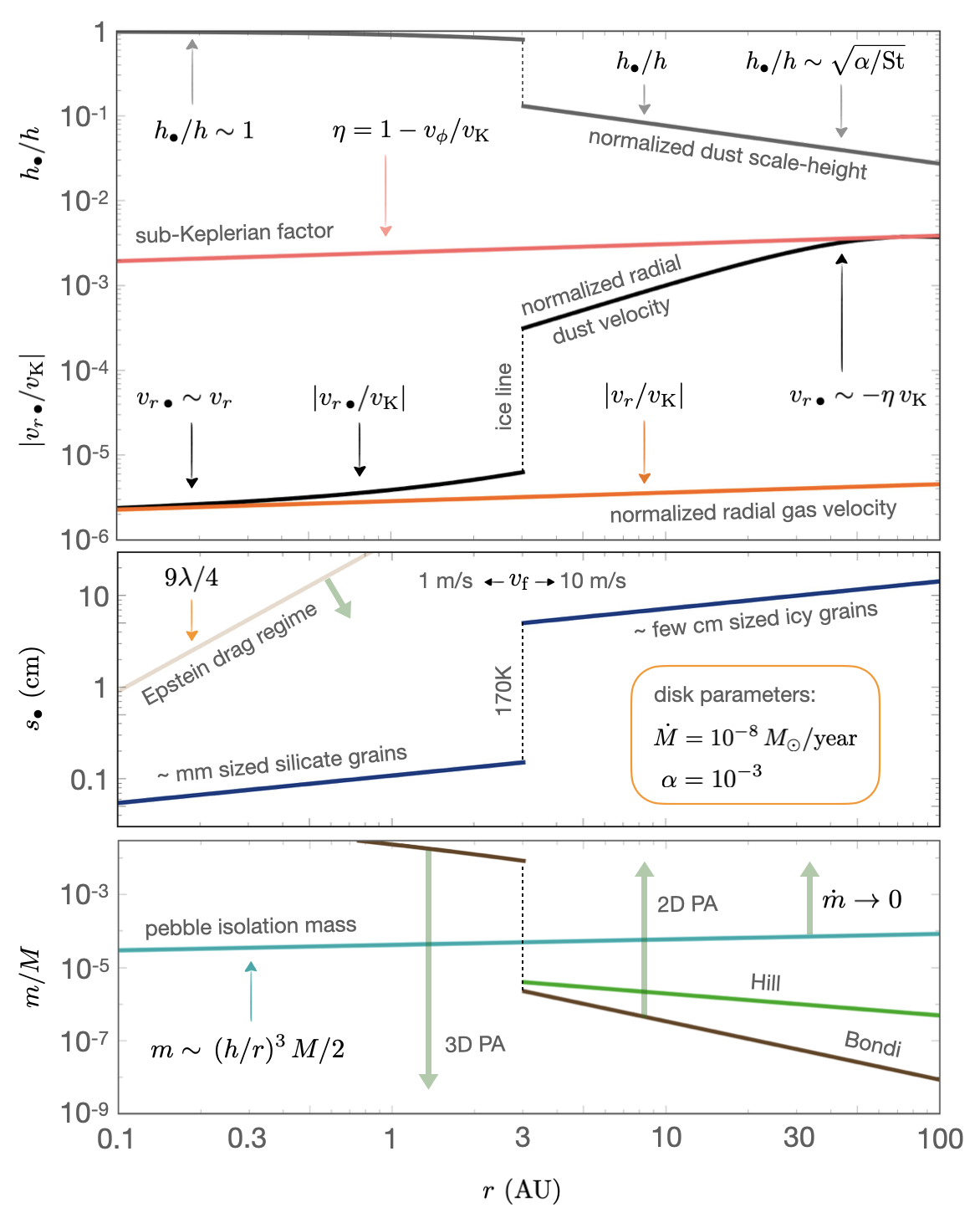}
\caption{Vertical thickness of the dust sub-disk, radial drift of solids, particle size profile, and pebble accretion regimes in a fiducial ($\alpha=10^{-3}$ and $\dot{M}\sim10^{-8}\,M_{\odot}/$year) model disk. Owing to strong dust-gas coupling in the inner disk, the solid layer is well-mixed (such that $h_{\bullet}\sim h$) and experiences slow ($v_{r\,\bullet}\sim v_r$) inward drift. In the outer disk, dust settling is much more efficient, and the radial velocity of icy grains achieves a value comparable to its upper limit of $v_{r\,\bullet}\sim-\eta\,v_{\rm{K}}$. The model further predicts that particle radii within the inner disk do not exceed the millimeter scale, while readily growing to sizes as large as a few centimeters beyond the ice line. The change in physical regimes of dust evolution across the water ice sublimation front translates to an important distinction in the modes of pebble accretion between the inner and outer regions of the nebula. While rapid 2D capture of solid dust can ensue in the outer disk for relatively low-mass protoplanetary embryos, the inner disk is fuly confined to the comparatively inefficient 3D regime of pebble accretion.}
\label{fig:fig4}
\end{figure}

Owing to a balance between turbulent stirring and gravitational settling, dust grains within the nebula form an embedded solid sub-disk with a scale-height $h_{\bullet}$. In general, the scale-height of the dust cannot exceed the scale-height of the gas, and it is convenient to consider the normalized quantity\footnote{As in equation (\ref{sigmadust}), in equation (\ref{hh}), we have assumed a turbulent Schmidt number of unity.} $h_{\bullet}/h$ \citep{1995Icar..114..237D}:
\begin{align}
\frac{h_{\bullet}}{h}=\frac{1}{\sqrt{1+\mathrm{St}/\alpha}}.
\label{hh}
\end{align}
A profile of $h_{\bullet}/h$ as a function of $r$ is shown as a gray curve in the top panel of Figure (\ref{fig:fig4}). Notice that in the inner disk, $\mathrm{St}\ll \alpha$, and $h_{\bullet}/h$ tends towards unity, implying a well-mixed particle layer. Conversely, in the outer disk, $\mathrm{St}\gg \alpha$, and the solid sub-disk settles to a normalized thickness of $h_{\bullet}/h\approx\sqrt{\alpha/\mathrm{St}}\lesssim0.1$. Notwithstanding the definitiveness of these results, it is important to keep in mind that real protoplanetary disks will always host particles with Stokes numbers that are substantially smaller than that given by equation (\ref{Stokesresult}) in some (presumably small) proportion. In other words, even in a physical regime where $h_{\bullet}/h\ll1$, very small (e.g., $\sim$micron-sized) grains will remain vertically well-mixed and continue to reside within the upper layers of protoplanetary nebulae \citep{2020MNRAS.494.5134L}.

The radial drift rate of solids $v_{r\,\bullet}$, normalized by the Keplerian orbital speed, $v_{\rm{K}}$, is given by the well-known expression \citep{1986Icar...67..375N}:
\begin{align}
\frac{v_{r\,\bullet}}{v_{\rm{K}}}=\frac{v_{r}-2\,\mathrm{St}\,\eta\,v_{\rm{K}}}{v_{\rm{K}}\,\big(1 + \mathrm{St}^2 \big)}.
\label{vr}
\end{align}
Importantly, this drift arises both from the accretionary flow of the gas (given by the first term in the numerator of the above expression), as well as from the head-wind experienced by the particles due to the sub-Keplerian flow of the gas. The latter effect depends on the magnitude of pressure support within the disk, and is controlled by the sub-Keplerian factor, $\eta$, which is approximately equal to the square of the disk's geometric aspect ratio \citep{Armitagebook}. More specifically, for the disk-model at hand it is straightforward to show that $\eta=(51/40)\,(h/r)^2\sim3\times10^{-3}$. The radial profile of $\eta$ is shown on the top panel of Figure (\ref{fig:fig4}) with a pink line.

The absolute value of the normalized radial dust velocity $|v_{r\,\bullet}/v_{\rm{K}}|$ is shown on the top panel of Figure (\ref{fig:fig4}) with a black curve. As with the scale-height of the solid sub-disk, the ice-line plays an important role in separating the particle drift regimes. In the inner disk, dust drifts very slowly relative to Keplerian orbital speed, and approaches the accretionary velocity of the gas (shown as an orange line). In the outer disk, particle drift due to headwind dominates and for $r\gtrsim40\,$AU, as the stokes number tends to unity, the radial velocity approaches its maximal limit of $|v_{r\,\bullet}/v_{\rm{K}}|\rightarrow\eta$.

While the Stokes number itself quantifies the strength of dust-gas coupling, it is important to keep in mind that fundamentally, $\rm{St}$ is set by the physical size of dust grains. Accordingly, we can employ the Epstein drag law\footnote{This drag-law is only appropriate for particles with radii smaller than or equal to $9/4$ times the mean free path of the gas molecules. We confirm the applicability of this regime a-posteriori.} to relate the Stokes number to the particle radius, $s_{\bullet}$:
\begin{align}
\mathrm{St}=\sqrt{\frac{\pi}{8}}\frac{\rho_{\bullet}}{\rho}\frac{s_\bullet}{c_{\rm{s}}}\,\Omega.
\label{Epstein}
\end{align}
For definitiveness, we adopt particle densities of $\rho_\bullet=1\,$g/cc and $3\,$g/cc for icy and rocky grains, respectively. The particle radius is shown as a function of stellocentric distance in the middle panel of Figure (\ref{fig:fig4}). Interior to the ice-line, rocky grains possess characteristic radii on the order of a millimeter. Beyond the ice-line, however, our model predicts that icy particles can achieve radii of a few centimeters, even reaching a decimeter for $r\gtrsim30\,$AU.

As a final result of this section, we consider the accretion of dust onto planetary objects, with an eye towards delineating the relevant physical regimes. Crudely speaking, accretion of solid dust ensues when planetary gravity can detach drifting particles from the background gaseous flow, causing the grain to spiral in towards the planet. Given a planetary mass $m$, the characteristic length-scale over which this can occur is the modified Bondi radius $\mathcal{R}_{\rm{B}}=\mathcal{G}\,m/(\eta\,v_{\rm{K}}^2)\sqrt{\mathrm{St}\,M\,\eta^3/m}$, where the multiplicative factor within the square root accounts for the strength of dust-gas coupling \citep{LJ12}.

The rate of mass accretion itself further depends on whether the particle layer forms a thin sub-disk or is vertically extended. In the former case, dust capture proceeds in a 2D regime at the rate $\dot{m}_{\rm{2D}}=2\,\Sigma\,\sqrt{2\,\mathcal{G}\,m\,r\,\eta\,\mathrm{St}}$. Conversely, in the case of 3D accretion, the rate of planetary mass growth has the form $\dot{m}_{\rm{3D}}=m\,\Sigma\,\mathrm{St}\,\sqrt{2\,\pi\,\mathcal{G}\,r^3/M}/h_{\bullet}$ \citep{OrmelReview2017}. Equating these two expressions, the criterion for transition from 2D to 3D pebble accretion can be formulated as a critical planetary-to-stellar mass ratio:
\begin{align}
\bigg(\frac{m}{M} \bigg)_{\rm{2D-3D}}\sim\frac{4\,\alpha\,\eta\,h^2}{\pi\,r^2\,\mathrm{St}\,(\mathrm{St+\alpha})}.
\label{PA2D3D}
\end{align}

The RHS of this expression can be readily computed from results delineated in the proceeding text, and the critical mass ratio is shown as a function of orbital radius in the bottom panel of Figure (\ref{fig:fig4}) with a brown line. Once again, the ice-line plays a decisive role: interior to the water-ice sublimation front, the pebble accretion process lies firmly in the 3D regime. Conversely, beyond the ice-line, 2D accretion is ensured for sufficiently massive proto-planets, with the critical mass ratio decreasing with increasing stellocentric distance. 

Beyond this 2D/3D boundary, there exists an additional transitionary mass scale, pertinent to two-dimensional accretion. In a regime where the Bondi accretion radius exceeds the Hill accretion radius $\mathcal{R}_{\rm{H}}=r\,(10\,\mathrm{St}\,m/(3\,M))^{1/3}$, the latter becomes the relevant pebble capture length-scale \citep{Ida2016}. This Bondi-Hill partition corresponds to a different critical mass scale, which can be computed by setting $\mathcal{R}_{\rm{H}}\sim\mathcal{R}_{\rm{B}}$:
\begin{align}
\bigg(\frac{m}{M} \bigg)_{\rm{B-H}}\sim\frac{100\,\eta^3}{9\,\mathrm{St}}.
\label{PA2D3D}
\end{align}
This relation is depicted as a dark green curve on the bottom panel of Figure (\ref{fig:fig4}), and shows that within the context of our disk model, the transition from Bondi to Hill accretion in the outer disk ensues when the planetary mass becomes comparable the Earth-Mars range, depending on the orbital period.

With these analytic results in hand, it is important to keep in mind that planetary accretion is not a process that is fixed to a given orbital radius. Rather, disk-driven orbital decay can cause growing planets to migrate across the ice line, effectively halting their accretion process. Conversely, rapid outward migration of protoplanetary embryos facilitated by the thermal torque \citep{2017MNRAS.472.4204M,2019MNRAS.486.5690G} can cause objects to enter a domain where bonafide 2D accretion can ensue. In other words, the interplay between accretion and migration constitutes an important driver of emergent diversity within planetary systems.

Irrespective of the exact regime in which pebble accretion operates, planetary growth cannot continue without bounds. Rather, \citet{Lambrechts2014} have shown that the pebble accretion process is self-limiting, and naturally terminates when a planet reaches the pebble ``isolation mass''. Notably, this mass-scale is related to the cube of the disk's geometric aspect ratio, and to within a factor of order unity is equal to the so-called ``thermal mass'' scale:
\begin{align}
\bigg(\frac{m}{M} \bigg)_{\rm{iso}}\sim\frac{1}{2}\bigg(\frac{h}{r} \bigg)^3.
\label{PA2D3D}
\end{align}
The pebble isolation mass is shown in the bottom panel of Figure (\ref{fig:fig4}) as a cyan line. 

Cumulatively, our simple analysis points to a clear ranking of pebble accretion regimes that is strikingly different in the inner and outer regions of the disk. Interior to the ice-line, pebble accretion proceeds entirely in the 3D regime, and is therefore comparatively inefficient. Exterior to the ice-line, the 3D regime of dust capture gives way to 2D Bondi accretion at low proto-planetary masses, and switches to 2D Hill accretion as the planetary mass increases. These results suggest that while planets may readily reach the isolation mass through pebble accretion in the outer disk, a different mode of growth (i.e., pairwise collisions among planetesimals) is likely to dominate within the inner regions of protoplanetary nebulae.

\section{Conclusion} \label{sec:conc}

In this work, we have outlined a simple model for the strength of dust-gas coupling within protoplanetary disks, as characterized by the Stokes number. The qualitative basis of our theoretical picture can be understood as follows. Through hydrodynamic drag, turbulent fluctuations within circumstellar nebulae excite the velocity dispersion among dust particles. In addition to being dependent upon the gas surface density and the magnitude of turbulent viscosity itself, the square of the dust velocity dispersion scales with Stokes number \citep{OrmelCuzzi2007}.

While particle coagulation leads to larger values of $\rm{St}$ and an enhanced spread of particle velocities, collisional fragmentation -- which is quantified by a composition-dependent threshold speed \citep{Blum1993} -- serves as the limiting mechanism for the growth of $\rm{St}$. Therefore, an equivalence between the dust velocity dispersion and the fragmentation speed yields an estimate for the maximum value of the Stokes number attainable by particles at a given orbital radius for a specified combination of disk parameters $\alpha$ and $\dot{M}$ (equation \ref{Stokesresult}; Figure \ref{fig:fig3}).

The key implications of our results are readily summarized. While the Stokes number exhibits an almost-linear dependence on the orbital radius, a much more dramatic (factor of $\sim10^{2}$) enhancement of $\rm{St}$ ensues across the water sublimation line, owing to an order-of-magnitude increase in the fragmentation velocity of the grains. For nominal disk parameters of $\alpha=10^{-3}$ and $\dot{M}=10^{-8}\,M_{\odot}/$year, the characteristic values of the Stokes number evaluate to $\mathrm{St}\sim10^{-4}$ (corresponding to a particle size of order a millimeter) and $\mathrm{St}\sim10^{-1}$ (corresponding to a particle size of a few centimeters) in the inner and outer regions of the disk, respectively. Because of this dichotomy, both vertical settling as well as radial drift of dust are likely to be efficient within the outer regions of protoplanetary disks\footnote{We note, however, that numerous mechanisms may act to trap dust locally, suppressing large-scale radial drift.}. Conversely, silicate dust within inner regions of protoplanetary disks is expected to be vertically well-mixed, and sink towards the host star at a rate comparable to the accretionary flow of the gas (see Figure \ref{fig:fig4}).

There exist a number of avenues in which our model can be expanded upon. Perhaps most trivially, our ``active-disk'' treatment of the energy balance can be extended to account for ``passive'' stellar irradiation as well as more realistic radiative transfer. Additionally, it is possible to consider radially variable accretion rates \citep{2012A&A...537A..61L} or incorporate non-viscous drivers of gas accretion, such as disk winds \citep{2021MNRAS.tmp.3115T}, although this would complicate the relationship between $\dot{M}$ and the surface density. As yet another example, the viscosity parameter itself can be endowed with a radial dependence \citep{Penna2013}, given that distinct modes of (magneto-)hydrodynamic turbulence are expected to operate in different regions of the disk \citep{2019PASP..131g2001L}. Similarly, a more sophisticated treatment of the fragmentation threshold can be readily employed \citep{2008ARA&A..46...21B,2020MNRAS.492..210V}, yielding additional radial structure within the Stokes number profile. Although each of these (and other) developments are undoubtedly worthwhile, it is unlikely that the principal results of our work will be altered dramatically. In other words, our determination that dust-gas coupling is taut in the inner disk and marginal beyond the ice-line is likely robust. 

A number of physical processes, including the onset of the streaming instability as well as the accretion of pebbles, exhibit a pronounced dependence on the Stokes number, with higher values of $\mathrm{St}$ generally corresponding to faster growth \citep{2017A&A...606A..80Y,Ida2016}. Despite the stark difference in characteristic Stokes number that ensues across the water-ice sublimation front, however, planetesimal formation is expected to operate efficiently in the inner and outer regions of the disk alike. In particular, assuming parameters similar to those derived herein (i.e., $s_{\bullet}\sim\,$mm in the inner disk, $s_{\bullet}\sim\,$cm in the inner disk,) recent work of \citet{2021NatAs.tmp..264M} has demonstrated how radial concentration of solids near the silicate and H$_2$O sublimation lines naturally leads to the contemporaneous generation of rocky and icy planetesimals through gravito-hydrodynamic instabilities (see also \citealt{2016A&A...594A.105D,2021NatAs.tmp..262I}).

Growth beyond the planetesimal stage, however, is a different matter. While our results indicate that capture of pebbles can drive rapid accretion of proto-planets beyond the ice-line \citep{LJ12}, the smallness of the Stokes number in the inner disk indicates that accretion of silicate pebbles is markedly inefficient. Therefore, pairwise collisions among planetesimals likely constitute the dominant process through which rocky planets coalesce \citep{2009ApJ...703.1131H}. This difference in growth modes stems primarily from the fact that pebble accretion in the inner disk is expected to proceed in the (slow) 3D regime, whereas vertical settling of grains in the outer disk allows for rapid accumulation of icy dust through 2D pebble accretion. Our results thus indicate that beyond constituting a partition in the chemical composition of solids, the ice-line represents a veritable divide in physical accretion regimes, and thus plays a decisive role in sculpting the large-scale architecture of planetary systems.


\paragraph{Acknowledgments.} We thank the anonymous referee for providing a thorough and insightful referee report. K. B. is grateful to Caltech, Observatoire de la C\^ote d'Azur, the David and Lucile Packard Foundation, and the National Science Foundation (grant number: AST 2109276) for their generous support. A. M. acknowledges support from the ERC advanced grant HolyEarth N. 101019380.

\end{document}